%Paper: hep-th/9506160
%From: atish@theory.caltech.edu (Atish Dabholkar)
%Date: Sun, 25 Jun 95 19:19:58 PDT

\input harvmac

\overfullrule=0pt
\def\Title#1#2{\rightline{#1}\ifx\answ\bigans\nopagenumbers\pageno0\vskip1in
\else\pageno1\vskip.8in\fi \centerline{\titlefont #2}\vskip .5in}

\font\ticp=cmcsc10

\font\secfont=cmcsc10

%
%defs
%
\baselineskip=18pt plus 2pt minus 2pt

\def\ajou#1&#2(#3){\ \sl#1\bf#2\rm(19#3)}
%
%Caligraphic
\def\CH{{\cal H}}

%
%\def\eightbig#1{{\hbox{$\textfont0=\ninerm\textfont2=\niesy\left#1\vbox
% to6.5pt{}
%\right.\n@space$}}}
%\eightbig{N}

%Roman

%

%Greek

\def\r{\rho}

\def\s{\sigma}

\def\d{\delta}
\def\m{\mu}
\def\n{\nu}
\def\p{\pi}
\def\e{\epsilon}

\def\l{{\lambda}}

\def\G{\Gamma}

%

%This paper

\def\TrH#1{ {\raise -.5em
                      \hbox{$\buildrel {\textstyle  {\rm Tr } }\over
{\scriptscriptstyle \CH _ {#1}}$}~}}

\def\IZ{\relax\ifmmode\mathchoice
{\hbox{\cmss Z\kern-.4em Z}}{\hbox{\cmss Z\kern-.4em Z}}
{\lower.9pt\hbox{\cmsss Z\kern-.4em Z}}
{\lower1.2pt\hbox{\cmsss Z\kern-.4em Z}}\else{\cmss Z\kern-.4em Z}\fi}
\def\IC{\relax\hbox{$\inbar\kern-.3em{\rm C}$}}
\def\IR{\relax{\rm I\kern-.18em R}}
\def\1{\relax 1 { \rm \kern-.35em I}}
\font\cmss=cmss10 \font\cmsss=cmss10 at 7pt

%

%Shortforms
\def\frac#1#2{{#1 \over #2}}

\def\p+{{\partial_+}}

\def\half{{1 \over 2}}

\def\apm{\alpha^{\prime}}

%refs

%

\Title{\vbox{\baselineskip12pt
\hbox{CALT-68-2002}
\hbox{\ticp doe research and}
\hbox{\ticp development report}
\hbox{}
\hbox{hep-th/9506160}
}}
{\vbox{\centerline {\bf TEN DIMENSIONAL HETEROTIC STRING AS A SOLITON} }}

\centerline{{\ticp Atish Dabholkar}}
\vskip.1in
\centerline{\it Lauritsen Laboratory of  High Energy Physics}
\centerline{\it California Institute of Technology}
\centerline{\it Pasadena, CA 91125, USA}
\centerline{ e-mail: atish@theory.caltech.edu}
\vskip .1in

\bigskip
\centerline{ABSTRACT}
\medskip

It is shown that the heterotic string emerges as a soliton in
the type I superstring theory in ten dimensions.
The collective coordinates of the soliton are described by a
smooth, chiral worldsheet theory. There are eight bosonic and
eight right-moving fermionic zero modes that arise from the
partially broken supertranslations. In addition, there are
496 charged bosonic zero modes of the gauge field that
describe a left-moving WZNW model on a $spin(32)/{Z_2}$ group
manifold. Small, stable loops of the solitonic string furnish
the massive states required by duality that transform as
spinors of $spin(32)$.

\bigskip

\bigskip
%\baselineskip=20pt plus 2pt minus 2pt
%\draft
\Date{June, 1995}

\vfill\eject

%References
\def\npb#1#2#3{{\sl Nucl. Phys.} {\bf B#1} (#2) #3}
\def\plb#1#2#3{{\sl Phys. Lett.} {\bf #1B} (#2) #3}
\def\prl#1#2#3{{\sl Phys. Rev. Lett. }{\bf #1} (#2) #3}
\def\prd#1#2#3{{\sl Phys. Rev. }{\bf D#1} (#2) #3}

\def\cmp#1#2#3{{\sl Comm. Math. Phys. }{\bf #1} (#2) #3}

%-------------------
% references
%-------------------
%

\lref\chsrev{C. G. Callan, J. A. Harvey and A. Strominger,
``Supersymmetric String
Solitons'' in String Theory and Quantum Gravity 1991: Proceedings of the
Trieste Spring School,
World Scientific, Singapore, 1991, hep-th/9112030.}
\lref\dabhharv{ A. Dabholkar and J. A. Harvey,  ``Nonrenormalization of
the Superstring Tension,''
\prl{63}{1989}{478}.}
\lref\dghr{ A. Dabholkar, G. Gibbons, J. A. Harvey and F. R. Ruiz,\hfill
``Superstrings and Solitons,'' \npb{340}{1990}{33}.}
\lref\witten{E. Witten, ``String Theory Dynamics in Various Dimensions,''
hep-th/9503124.}
\lref\wbran{C. G. Callan, J. A. Harvey and A. Strominger,
``Worldbrane Actions for String Solitons,'' \npb{367}{1991}{60}.}
\lref\hulltown{C.M. Hull and P.K.  Townsend,
``Unity of Superstring Dualities,'' QMW-94-30, R/94/33,
hep-th/9410167. }
\lref\strominger{ A. Strominger,  ``Superstrings with Torsion,''
\npb{274}{1986}{253}.}
\lref\duffkhur{M. Duff, preprint CTP-TAMU-49/94, hep-th/9501030\semi
M. Duff and R. Khuri, \npb{411}{1994}{473}, hep-th/9305142).}
\lref\harvstro{ J. A. Harvey and A. Strominger, ``The Heterotic String
is a Soliton,'' EFI-95-16, hep-th/9504047.}
\lref\senone{ A. Sen, ``String-String Duality Conjecture in Six
Dimensions and Charged Solitonic Strings,''
TIFR-TH-95-16, hep-th/9504027.}
\lref\gsw{M. B. Green, J. H. Schwarz, and E. Witten,
{\it Superstring Theory},  {\rm Vols. I and II} ,
Cambridge University Press (1987).}
\lref\mtw{C. W. Misner, K. S. Thorne, and J. A. Wheeler,
{\it Gravitation}, W. H. Freedman and Company (1973).}
\lref\ghs{D. Garfinkle, G. Horowitz and A. Strominger,
``Charged Black Holes in String Theory,'' \prd{43}{1991}{3140}.}
\lref\gepnwitt{D. Gepner and E. Witten, `` String Theory on
Group Manifolds,'' \npb{278}{1986}{493}.}
\lref\duffetal{M. Duff, G. Gibbons and P. Townsend, DAMTP/R-93/5,
hep-th/94-5124\semi
For a review see M. J. Duff,  R. R. Khuri, and J. X. Lu, ``String
Solitons,'' hep-th/9412184.}
\lref\sentwo{A. Sen, ``Macroscopic Charged Heterotic String,'' hep-th/9206016,
\npb{388}{1992}{457}\semi
S. Hassan and A. Sen, ``Twisting Classical Solutions in
Heterotic String Theory,'' \npb{375}{1992}{103}.}
\lref\chstwo{C. Callan, J. Harvey and A. Strominger,
``Worldsheet approach to heterotic solitons and instantons,''
\npb{359}{1991}{611}.}
\lref\hughpolc{J. Hughes and J. Polchinski, \npb{278}{1986}{147}.}
\lref\wittoliv{E. Witten and D. Olive, \plb{78}{1978}{97}.}
\lref\osbborne{H. Osborne,
``Topological Charges for N=4 Supersymmetric
Gauge Theories and Monopoles of Spin 1,''
\plb{83}{1979}{321}.}
\lref\montoliv{C. Montonen and D. Olive, \plb{72}{1977}{117}\semi
P. Goddard, J. Nyuts and D. Olive, \npb{125}{1977}{1}.}
\lref\wznw{E. Witten, ``Non-Abelian Bosonization in Two Dimensions,''
\cmp{92}{1984}{451}.}
\lref\swone{N. Seiberg and E. Witten, ``Electromagnetic Duality,
Monopole Condensation and Confinement in N=2 Supersymmetric
Yang-Mills Theory,'' hep-th/9407087.}
\lref\stroming{A. Strominger, ``Heterotic Solitons,''
\npb{343}{1990}{167}.}
\lref\senthree{A. Sen, \plb{329}{1994}{217}, hep-th/9402032.}

\newsec{Introduction}

It has recently been conjectured that in ten dimensions,
the type I superstring and the heterotic string with gauge
group $SO(32)$ are dual
to each other \witten . While the evidence for this duality
has so far been only at the level of the low energy effective
lagrangian, it is likely that the two string theories are exactly
dual to each other. One requirement of such an exact duality is
that the spectrum of BPS saturated states must match
in the two theories.
Of particular interest are the neutrally stable,
macroscopic winding states that exist in
the perturbative spectrum of the heterotic string
\refs{\dabhharv ,\dghr}. In the type I theory such states do not
occur perturbatively and hence must arise as solitons.
In this paper, we show that the macroscopic heterotic string does
indeed exist as a soliton in the type I theory. The collective
coordinates of this soliton are described by a smooth
worldsheet theory that is identical to
the worldsheet of the perturbative heterotic string.
Remarkably, the intricate chiral structure
of the heterotic string emerges from a theory of
unoriented strings.

A similar exact duality has been conjectured in six dimensions
between a heterotic string compactified on a four-torus and the
type IIA string compactified on a $K3$ surface
\refs{\hulltown ,\witten , \duffkhur}.
The corresponding heterotic soliton that arises in the type IIA
theory has been described in \refs{\harvstro ,\senone}.
At a generic point in the
$K3$ moduli space, the gauge symmetry is abelian and
the zero modes of the soliton are chiral worldsheet
bosons that live on an even, self-dual lattice with signature $(20, 4)$.
The current algebra is thus realized in the free boson representation.
In ten dimensions, the situation is quite different.
The spacetime coordinates and the right-moving fermions
are still realized as free fields.
The left-moving charge carrying current algebra, however, is
realized as a chiral WZNW model on the soliton worldsheet.
In six dimensions also,
at points of enhanced, nonabelian symmetry in the $K3$ moduli space,
the charged current algebra will be realized in this manner.

An important requirement for obtaining the heterotic string is
that the WZNW model be formulated not on
the simply connected group manifold of $spin(32)$ but rather on the
group manifold of $spin(32)/Z_2$.
For the perturbative heterotic string it follows from
modular invariance.
For the soliton, we can partially understand this from
the fact that the zero modes naturally give rise to
only left-moving currents. In order to describe them it
is necessary to obtain a WZNW model that factorizes holomorphically.
This essentially determines the correct identification.

Another requirement for consistency is that we have a level-one
current algebra. It is not apparent how to derive this condition
for the soliton, but it may conceivably follow from some topological
considerations. In
what follows, we shall simply assume
this for nonperturbative consistency.

In the next section we describe the heterotic string soliton and
the structure of zero modes in its background.
In the last section we briefly comment upon
the spacetime structure of the soliton and the relation
of our work to others in different dimensions.

\newsec{Heterotic String as a Soliton in the Type I String Theory}

The massless bosonic fields of the type I string theory consist of
the dilaton $\phi$ and the metric tensor $g_{MN}$ from the
NS-NS sector, an antisymmetric tensor $B_{MN}$ from
the R-R sector and an $SO(32)$ gauge field $A_M$
from the open string sector.
The low-energy action for these fields \refs{\gsw , \witten} is
\eqn\action{S = {1 \over {\apm}^4}\int d^{10}x \sqrt{-g} \left(
                e^{-2 \phi}(R + 4(\nabla \phi)^2)
                -{\apm \over 30} e^{-\phi}{\rm Tr} F^2
                 - {1 \over 3} H^2 \right),}
where $F=dA + A \wedge A$ and
the three-form antisymmetric tensor field strength is
given by
\eqn\anomaly{H=dB +\apm\left(\omega^L_3(\Omega_-)-{1\over 30}
			\omega^{YM}_3(A)\right)+\ldots \, .}
Here $\omega_3$ is the Chern-Simons three-form and
the connection
$\Omega_\pm$ is a non-Riemannian
connection related to the usual spin connection $\omega$ by
\eqn\connection{ \Omega_{\pm M}^{AB} = \omega_M^{~AB} \pm  H_M^{~AB}
+ (\half e^{NA} e^B_M \partial_N \phi - A \leftrightarrow B). }
When the action is written this way, the gauge fields have standard,
dilaton independent transformation laws $\d A= D\e$ and $\d B = d\l$,
but now the lagrangian no longer scales with an overall factor
of $e^{-2\phi}$. This is what we expect for the type I theory
where as usual $e^{2 \phi}$ is the loop counting parameter
and the $F^2$ term is higher order because it comes from the disc diagram.
The duality transformation for these fields is
$\phi ' = -\phi $, $g' = e^{-\phi} g$, $A'=A$ and $B' =B$
where the unprimed fields are type I and the primed fields are heterotic.
Under duality one obtains the heterotic string Lagrangian
with the normalization used in \chsrev .

For type I string theory compactified on a very
large circle, we expect
to find macroscopic string solitons that
wrap around this circle. We can take $\s^\mu (\mu = 0,1)$
as the coordinates on the worldsheet of the soliton and
$y^m (m=2,...,9)$ as the transverse coordinates.
The solitonic string solution is then given by
\eqn\solution{\eqalign{
{g_{\mu\nu}} &= {e^{-\phi}}{\eta_{\mu\nu}}.\cr
{g_{mn}}     &= {e^{+\phi}}{\delta_{mn}}.\cr
{H}_{\mu\nu p}
  &= {-\half}{\epsilon_{\mu\nu}\partial_p e^{-2\phi} },\cr
{e^{2\phi}}  &=    {e^{2\phi_0}}  + {Q \over y^6}, \cr}}
where $y$ is the radial distance in the transverse
space, $\eta_{\m\n}$ is the flat lorentzian metric on the
worldsheet and $\e_{\m\n}$
is the antisymmetric tensor with $\e_{01} = - \eta_{00}= 1$.
The parameter $Q$ is the axion charge per unit length:
\eqn\charge{ Q =  2 \int_{S^7} e^{-\phi}{}^* H .}
It is this Ramond-Ramond charge that gives the soliton string
its orientation. The original type I string is unoriented because
the perturbative string worldsheet does not couple to
the $B_{MN}$ field from the RR sector.
It is easy to see that the solution saturates a
Bogomol'nyi bound \refs{\dghr} and the mass per unit
length equals the charge $Q$. As a result,
half the supersymmetries remain unbroken in the
soliton background.

Let us now discuss the various zero modes. There are
eight bosonic zero modes coming from the broken
translational invariance that correspond to the location
of the soliton. We denote these
by $X^m (\s )$ which are free fields on the worldsheet.
Then there are eight right-moving, fermionic zero
modes coming from the broken supersymmetries.
The supersymmetry transformations for the fermions are
\eqn\susyvar{ \eqalign{\delta \lambda & = (\G^{M} \partial_M \phi -
                       {1 \over 6} H_{MNP}\G^{MNP}) \epsilon  \cr
                        \delta \psi_M & = (\partial_M +
                                           {1 \over 4}\Omega_{-M}^{AB}
                                          \G_{AB}) \epsilon,   \cr }}
All spinors are Majorana-Weyl in ten dimensions. The gravitino $\psi_M$
and the supersymmetry variation parameter $\e$ have positive chirality
whereas the dilatino $\l$ has negative chirality. Under the natural
embedding $SO(1,9) \rightarrow SO(1,1) \times SO(8)$ the spinor
$\e$ decomposes as
\eqn\spinor{\e \rightarrow 8^+_+ + 8^-_- ,} where the
superscript and the subscript refer to $SO(8)$ and $SO(1,1)$
chiralities respectively. Half of these supersymmetries
$8^-_-$ are annihilated by the solution \solution .
The unbroken supersymmetries imply worldsheet supersymmetry
\hughpolc . The other half $8^+_+$ gives rise to right-moving
zero modes on the soliton worldsheet
which we denote by $ S^a_+(\s )$. These are
free, right-moving Majorana fermions on the worldsheet where $a$ is
the $SO(8)$ spinor index with positive chirality.

As in \refs{\dghr ,\stroming ,\chstwo},
the effective worldsheet action for these zero modes
can be easily obtained. The bosonic part of the action is given by
\eqn\GS{ {N \apm \over \kappa}
\int d^2\sigma e^{-\phi}( \eta^{\m\n}\partial_\m
X^m\partial_\n X^n
g_{mn})+ \e^{\m\n}\partial_\m X^m\partial_\n X^n
B_{mn},}
where $N$ is a normalization factor, $\kappa$ is
the gravitational coupling  and $B$
is the RR, Kalb-Ramond field.
Using heterotic variables and
the fact that the heterotic string tension
is ${\kappa \over \apm}$,
one obtains the usual action
in static gauge ($X^\mu \sim \sigma^\mu$).
The action for the fermionic zero modes is determined
by worldsheet supersymmetry \hughpolc\ and
is similar to the light-cone Green-Schwarz action for
the field $S^a_+$ \gsw .

We now turn to the charged zero modes that come from
solving the equation of motion for the gauge fields.
In the differential form notation, the relevant part of the action
\action\ is
\eqn\actiontwo{
- {2\over {\apm}^4}\int
                ( {1 \over 3} H \wedge * H \, + \,
                {\apm \over 30} e^{-\phi}{\rm Tr} F \wedge * F ),
}
where $*$ is the Hodge star operation with the conventions
of \mtw . In varying with respect to $A$ we have to remember
that $H$ depends on $A$ through the Yang-Mills Chern-Simons
three form \anomaly . Since
$\d w^3_{YM} = 2\, {\rm Tr}\, \d A \wedge F + d (A\wedge \d A )$, and
$d*H=0$ is one of the equations of motion, we get,
\eqn\chargeone{
D * (e^{-\phi} F ) = -2 F \wedge * H .}
We would like to solve this equation for small fluctuations of
the gauge field in the background of the remaining fields given by
solution \solution .
Now recall that $ H = -\half \e \wedge d e^{-\phi}$ and
$* H = \half e^{\phi} {\hat *}  d e^{-2\phi}$ where ${\hat *}$ is
the Hodge star operation either in the transverse space or along
the worldsheet depending on whether the forms are cotangent to
the transverse space or the worldsheet. The zero mode equation
can then be solved by the ansatz
$A =  a(\s )de^{-2\phi}$ if,
\eqn\leftcondition{\eqalign{
    da &= {\hat *} da , \cr
    d^{\hat *}d a &= 0.\cr}}
Since $A$ is a one-form valued in the Lie algebra
of $SO(32)$, we obtain $496$ bosonic left-moving
zero modes. The chirality of the zero modes is correlated with
the sign of the exponent in the ansatz. The right-moving zero
modes are proportional to $d e^{2\phi}$ and are not normalizable.
Note that $A\wedge A = 0$ for our ansatz  and for constant $a$
the zero modes are pure gauge.

The dimension of the moduli space of the left-moving
charge-carrying modes can also be seen by a slightly different
argument.
The Lagrangian in \action\ is invariant under a global group
$SO(32)$ that is generated by gauge transformations with a gauge
parameter that is constant at infinity. This group leaves
the neutral solution \solution\ invariant and does not generate
a new solution. However, one can use the solution-generating
technique of Sen \sentwo , to obtain a solution that carries charge
under a $U(1)$. This is achieved by taking a $U(1)$ inside $SO(32)$
and embedding it inside the solution generating T duality group.
Once we have a solution that is charged under a $U(1)$
we can act with the $SO(32)$ to obtain a $496$ parameter
family of solutions.

We can write $a(\s) = \e^a(\s ) T_a$  where $T_a$
are the (anti-Hermitian) generators
in some representation. These zero modes then
parametrize, up to global identifications, the
group manifold of $spin(32)$. If we write
$g = e^{\e^a T_a}$ then
the field strength is given by,
\eqn\leftmovers{F = d\s^- \partial_{-}g g^{-1}\wedge de^{-2\phi} }
with $\partial_{+} (\partial_{-}g g^{-1}) =0$.
These equations of motion imply a chiral WZNW action \wznw\ along
the worldsheet.

We have taken the moduli to parametrize the universal covering
group but there can be discrete moddings by the center of
$spin(32)$.
In fact, such discrete
identifications are necessary since we would like to define the chiral
WZNW model as the holomorphic part of the full WZNW model.
In general, the WZNW model does not completely factorize.
For example, the level-one current algebra of $spin(32)$
contains four representations: singlet, vector, spinor
and conjugate spinor.  The one-loop partition function
picks the diagonal modular invariant and
is a sum of four terms \gepnwitt . Each term is an absolute  square of a
holomorphic term but the sum itself is not an absolute square.
However, the WZNW model on
$spin(32)/{Z_2}$ group manifold does give
a partition function that factorizes holomorphically.
Thus, the fact that we have only left-moving currents
essentially determines the correct global
identification on the group manifold.
This argument requires some knowledge about higher loop quantization
and it would be nice to obtain this identification completely
from semiclassical reasoning, if possible.
Under this $Z_2$ the spinor and the adjoint are invariant
whereas the vector and the conjugate spinor are not.
As a result, the spectrum obtained from the quantization of collective
coordinates will contain representations whose weights
are in the conjugacy classes of the adjoint and the spinor and lie
on an even, self-dual lattice of $spin(32)/{Z_2}$.

This completes the analysis of charged zero modes.
Together with \GS\ we have
recovered the entire structure of the ten-dimensional heterotic
string on the soliton worldsheet.

\newsec{Discussion}

We would now like to comment upon the singularity structure
of the soliton. All the zero modes that we have obtained are smooth
and normalizable and thus describe a smooth worldsheet. The antisymmetric
field-strengths $H$ and $F$ are also smooth everywhere. The dilaton
and the metric, on the other hand, appear to diverge at $y=0$.
The line element for the solution \solution\ is
\eqn\lineone{
ds^2 = e^{-\phi} (-dt^2 + dz^2) + e^{\phi} (dy^2 + y^2 d\Omega ^2), }
where $d\Omega^2$ is the volume element for the transverse
seven spheres. At $y=0$, $e^{\phi}$ diverges as ${1\over {y^3}}$.
If we change variables to $\rho \sim {1\over{y^{\half}}}$ then
the line element near $y=0$ becomes
\eqn\linetwo{
ds^2 = {A\over {\r^6}}(-dt^2 + dz^2) + d\r^2 + B \r^2 d\Omega ^2, }
where $A$ and $B$ are numerical constants. Along the surface
of constant $t$ and $z$ we see that singularity at $y=0$ has
receded an infinite geodesic distance away to $\rho = \infty$.
On this slice, the space has the geometry of a wormhole with
an infinite throat much like the geometry of some extremal black holes
\ghs .
One difference here is that as we go down the throat to
$\r \rightarrow \infty$, the circumference of the throat increases.
The line element in $t, z$ plane still seems singular and it is
not clear if the singularity is an infinite distance away if
we consider different approaches to $y=0$. In particular, it is not
obvious if the spacetime is null-geodesically complete.
The situation is somewhat better in six dimensions
\refs{\senone , \harvstro},
where the $t, z$ plane is flat and the singularity is manifestly
infinitely far away \duffetal .
In both dimensions, however, the dilaton grows down the throat as
we approch $y=0$ and the theory gets into strong coupling. This means
that no matter how small the coupling at asymptotic infinity, the
semiclassical approximation itself breaks down at a finite distance
down the throat once the coupling becomes of order one.
It appears, therefore, that both in ten and in six
dimensions, we lack an adequate criterion for
deciding which solutions should be regarded as nonsingular.

Another issue concerns the $\apm$ corrections to the leading
order solution. In six dimensions it was possible to argue that
the $\apm$ correction will vanish from $(4, 4)$ worldsheet
supersymmetry \refs{\harvstro ,\chstwo}. In ten dimensions, the
corrections appear to be nonzero. For example, the
generalized spin connection does not equal the gauge connection
for the charged solution. From the anomaly equation,
\eqn\curlh{dH=\apm (trR \wedge R- {1 \over 30}Tr F \wedge F), }
we see that $H$ would have to be modified at the next order.
We will have a left-right asymmetric solution and it seems unlikely
that worldsheet supersymmetry will protect the lowest order solution.

Even though the solution will be modified at higher orders,
it is reasonable to assume that the structure of zero modes
will be essentially unaltered. This is because the zero modes
are normalizable and in fact behave as ${y^6}$ near $y=0$.
Thus, they are well localized near the throat, far away from the
singularity. The $\apm$ expansion is an expansion in $\apm/{y^2}$,
therefore, the zero modes will not be substantially modified at large $y$.
Unless the $\apm$ corrections create new singularities near $y=0$
for these zero modes, they will remain normalizable.
Further analysis of the corrected solution, its causal
structure and the zero modes certainly deserves further investigation.

So far we have considered macroscopic string solitons wrapped around
a large circle which are stable because of the winding.
But one can also contemplate small, wiggling loops
of the solitons. Some of these solitons, if they exist,
will also be stable. In particular,
a soliton that carries the smallest spinor charge of $spin(32)$ will
be absolutely stable. The states in the
perturbative spectrum of the type I string carry only integer charges,
therefore a soliton with half-integer charge cannot decay into them.
These small, stable
loops of solitons will provide the massive states
with spinor charges that are required by duality.

It may seem puzzling that in \refs{\senone ,\harvstro}
the number of charged zero modes equals the rank of the group,
whereas our analysis in six dimensions will give as many
zero modes as the dimension of the group. This happens because
at a generic point in the $K3$ moduli space of type IIA
compactifications, where the gauge group is completely abelian,
the rank equals the dimension of the group. At special points of enhanced
symmetry, extra massless states appear and the low energy
Lagrangian changes abruptly. With the new low-energy Lagrangian
at these points, one will find that the number of zero
modes also jumps abruptly in accordance with the structure that
we have discussed.
Thus, the emergence of a chiral, WZNW model on a group
manifold with appropriate discrete identifications is a general
property of these solitons applicable also in six dimensions
at various points of nonabelian symmetry.

The results of this paper in ten dimensions,
along with the results of \refs{\senone , \harvstro} in six
dimensions provide a satisfying picture of `exact'
string-string duality. This duality is analogous to the exact
duality for $N=4$ supersymmetric gauge theories in four dimensions
\refs{\montoliv , \wittoliv , \osbborne ,\senthree}.
In field theory, the spectrum of BPS saturated soliton states
is also important in the understanding of the
`effective' duality in $N=2$ theories
\refs{\swone}.
We expect that the soliton strings that we have described will
play a similar role in the understanding of effective string-string
dualities.
\vfill
\eject
\bigskip
\leftline{ \secfont Acknowledgements}
\bigskip
I would like to thank J.~Gauntlett and A.~Sen for useful discussions,
and J. H. Schwarz for a critical reading of the manuscript.
This work was supported in part by the U. S. Department of Energy
under Grant No. DE-FG03-92-ER40701.
\vfill
\eject

\listrefs
\end